\documentclass{emulateapj}
\usepackage{apjfonts}
\usepackage[]{natbib}
\usepackage{graphics}
     
\newcommand{\etal}{et al.}  
\newcommand{\per}{\ensuremath{^{-1}}}

\newcommand{\hal}{H\ensuremath{\alpha}}
\newcommand{\hbeta}{H\ensuremath{\beta}} 
\newcommand{\hst}{\emph{HST}}
\newcommand{\msun}{\ensuremath{M_{\odot}}}
\newcommand{\kms}{km s\ensuremath{^{-1}}}

\newcommand{\mbh}{\ensuremath{M_\mathrm{BH}}}

\newcommand{\msigma}{\ensuremath{\mbh-\sigma}}
\newcommand{\rblr}{\ensuremath{r_{\mathrm{BLR}}}}

\newcommand{\mgb}{Mg\ensuremath{b}}

\slugcomment{To appear in The Astrophysical Journal Letters}
\shorttitle{LOW-MASS END OF THE \msigma\ RELATION} 
\shortauthors{BARTH ET AL.}

\begin{document} 

\title{Dwarf Seyfert 1 Nuclei and the Low-Mass End of the \msigma\ Relation}

\author{Aaron J. Barth\altaffilmark{1}, Jenny
  E. Greene\altaffilmark{2}, and Luis C. Ho\altaffilmark{3}}

\altaffiltext{1}{Department of Physics and Astronomy, 4129 Frederick
  Reines Hall, University of California, Irvine, CA 92697-4575;
  barth@uci.edu}

\altaffiltext{2}{Harvard-Smithsonian Center for Astrophysics, 60
  Garden St., Cambridge, MA 02138; jgreene@cfa.harvard.edu}

\altaffiltext{3}{The Observatories of the Carnegie Institution of
  Washington, 813 Santa Barbara Street, Pasadena, CA 91101;
  lho@ociw.edu}

\begin{abstract}

To examine the relationship between black hole mass and host galaxy
velocity dispersion for low black hole masses, we have measured the
velocity dispersions of 15 Seyfert 1 galaxies from the catalog of
Greene \& Ho (2004).  These Seyferts were selected from the Sloan
Digital Sky Survey to have estimated black hole masses below $10^6$
\msun.   The data are consistent with a straightforward downward
extrapolation of the local \msigma\ relation, indicating that this
correlation extends over a range of more than four orders of magnitude in
black hole mass.  The rms scatter of the sample about the extrapolated
\msigma\ relation is 0.57 dex, consistent with the expected scatter of
single-epoch mass estimates for Seyfert 1 galaxies.

\end{abstract}


\section{Introduction}

The correlation between black hole mass and stellar velocity
dispersion in nearby galaxies (the \msigma\ relation) is the
foundation upon which our understanding of supermassive black hole
demographics is built \citep{fm00, geb00a, tre02}.  To date, however,
our knowledge of the \msigma\ relation is largely limited to galaxies
with massive bulges, velocity dispersions of $\sim70-400$ \kms, and
$\mbh > 3\times10^6$ \msun.  Elucidating the demographics of
lower-mass black holes, below $10^6$ \msun, can address key questions
regarding the formation environments of massive black holes: for
example, is there a minimum galaxy mass or velocity dispersion below
which black holes are unable to form or grow \citep{bsf04}?  Are black
holes easily ejected from dwarf galaxies by gravitational recoil
following black hole mergers \citep{mer04}?  Extending the local black
hole mass function to masses below $10^6$ \msun\ would also aid in
prediction of event rates for black hole mergers detectable by
\emph{LISA} \citep[e.g.,][]{wl03}.

If black holes do exist in the centers of dwarf galaxies and very
late-type spirals, then the subset that are undergoing accretion can
potentially be detected as active galactic nuclei (AGNs).  However,
very few AGNs have been identified with black hole masses below $10^6$
\msun\ or in host galaxies with $\sigma < 60$ \kms.  These include NGC
4395 \citep{kra99, fh03}, POX 52 \citep{bar04}, possibly NGC 4051
\citep{she03}, and a few other narrow-line Seyfert 1 galaxies
\citep{bot04b, gm04}.  The first systematic survey for AGNs with
low-mass black holes was recently carried out by \citet{gh04}
(hereinafter GH), who examined the Sloan Digital Sky Survey (SDSS)
Data Release One (DR1) archives to find 19 Seyfert 1 galaxies having
estimated black hole masses below $10^6$ \msun.

The stellar velocity dispersions for the GH sample cannot be
determined from the SDSS spectra, because the instrumental dispersion
of $\sim70$ \kms\ sets a practical lower limit to the range of
velocity dispersions that can be measured, and in most cases the
spectra are dominated by AGN emission.  Here, we present new
measurements of the stellar velocity dispersions for 13 galaxies from
this sample and two additional Seyferts we identified from SDSS DR2
meeting the same selection criteria: SDSS J080907.58+441641.4 and SDSS
J215658.30+110343.1.

\section{Data and Measurements}

\begin{deluxetable*}{lccccccccc}
\tablewidth{6.8in} \tablecaption{Observations and Measurements}
\tablehead{ \colhead{Object} & \colhead{$z$} & \colhead{Exposure} &
\colhead{S/N} & \colhead{Line Strength} & \colhead{$\sigma$(\mgb)} &
\colhead{$\sigma$(Ca II)} & \colhead{FWHM(\hal)} &
\colhead{$\log(\lambda L_\lambda$)} & \colhead{$\log(\mbh/\msun)$} \nl
\colhead{} & \colhead{} & \colhead{(s)} & \colhead{} & \colhead{} &
\colhead{(\kms)} & \colhead{(\kms)} & \colhead{(\kms)} &
\colhead{(5100 \AA)} & \colhead{} }

\startdata
GH01 & 0.077 & 2982 & 17 & $0.26\pm0.08$ & $36\pm6$ & $34\pm8$ & 950  & 43.1 & 6.28 \\
GH02 & 0.030 & 6000 & 21 & $0.52\pm0.05$ & $50\pm4$ & $59\pm8$ & 690  & 41.9 & 5.16 \\
GH03 & 0.102 & 3000 & 10 & $0.45\pm0.07$ & $55\pm6$ & \nodata  & 820  & 42.9 & 6.01 \\
GH04 & 0.043 & 2700 & 32 & $0.27\pm0.05$ & $60\pm4$ & $57\pm7$ & 760  & 42.8 & 5.87 \\
GH05 & 0.074 & 3600 & 26 & \nodata       & \nodata  & \nodata  & 720  & 43.1 & 6.04 \\
GH06 & 0.100 & 1800 & 10 & $0.36\pm0.06$ & $48\pm7$ & \nodata  & 1240 & 42.9 & 6.37 \\
GH07 & 0.094 & 3600 & 17 & $0.17\pm0.04$ & $39\pm6$ & \nodata  & 1140 & 43.0 & 6.37 \\
GH10 & 0.081 & 1800 & 29 & $0.17\pm0.04$ & $49\pm7$ & $54\pm8$ & 720  & 43.3 & 6.18 \\
GH11 & 0.081 & 1800 & 19 & $0.37\pm0.08$ & $64\pm6$ & $62\pm7$ & 2110 & 43.2 & 7.04 \\
GH13 & 0.126 & 1800 & 12 & \nodata       & \nodata  & \nodata  & 980  & 43.3 & 6.45 \\
GH14 & 0.028 & 1800 & 14 & $0.34\pm0.05$ & $48\pm7$ & $59\pm7$ & 770  & 41.9 & 5.26 \\
GH16 & 0.069 & 1100 & 19 & $0.32\pm0.07$ & $69\pm6$ & $83\pm8$ & 1170 & 42.5 & 6.04 \\
GH17 & 0.099 & 3500 & 20 & $0.27\pm0.06$ & $58\pm7$ & \nodata  & 930  & 43.1 & 6.26 \\
GH18 & 0.183 & 3600 & 12 & $0.49\pm0.17$ & $81\pm9$ & \nodata  & 1770 & 43.1 & 6.81 \\
GH19 & 0.036 & 3600 & 20 & $0.69\pm0.19$ & $58\pm4$ & $53\pm4$ & 1890 & 41.7 & 5.89 \\
0809+4416 & 0.054 & 3600 & 22 & $0.47\pm0.11$ & $65\pm4$ & \nodata & 1100 & 42.8 & 6.19 \\
2156+1103 & 0.108 & 3600 & 34 & $0.23\pm0.06$ & $81\pm8$ & \nodata & 780 & 43.5 & \phn6.39 
\enddata

\tablecomments{Object names are those given in the \citet{gh04}
  catalog.  S/N is the mean signal-to-noise ratio per pixel in the ESI
  spectrum in the spectral region around \mgb\ used to measure
  $\sigma$. ``Line Strength'' gives the strength of stellar absorption
  features in the AGN spectrum relative to those in the template
  stars, over the blue fit region.  FWHM(\hal) is the width of the
  broad component of \hal, as determined by multi-Gaussian model fits.
  The AGN continuum luminosity $\log[\lambda L_\lambda$(5100 \AA)] is
  from \citet{gh04}; $\lambda L_\lambda$ is in units of erg s\per.
  The luminosities are corrected for Galactic extinction and are
  calculated for a cosmology with $H_0=72$ km s\per\ Mpc\per,
  $\Omega_m=0.3$, and $\Omega_\Lambda=0.7$.}
\label{table1}
\end{deluxetable*}

\subsection{Observations and Data Reduction}

The observations were obtained during the nights of 2003 November 24,
2004 February 15--17, and 2004 October 9--10 UT at the Keck II
telescope, using the ESI spectrograph \citep{she02}.  A 0\farcs75 slit
width was used, giving an instrumental dispersion of $\sigma_i \approx
22$ \kms.  The spectra cover the wavelength range 3800--10900 \AA\
across ten echelle orders, at a uniform scale of 11.5 \kms\ pixel\per.
For each observation, the slit was oriented at the parallactic angle.
Flux standards and red giant stars (G8III--K5III) for use as velocity
templates were observed each night during twilight.  Spectra were
extracted with a 1\arcsec\ extraction width, and wavelength- and
flux-calibrated using standard techniques.

\subsection{Stellar Velocity Dispersions}

We measured the stellar kinematics by fitting broadened stellar
spectra to the galaxy spectra, following techniques described by
\citet{bhs02}.  A Gaussian velocity broadening function was assumed,
and the fits included both an additive featureless continuum component
and a multiplicative low-order polynomial that can compensate for
reddening in the galaxy spectrum.  Fits were performed in two spectral
regions: $\sim5050-5400$ \AA, which includes \mgb, Fe5270, and
numerous weaker features; and $\sim8450-8700$ \AA, covering the
\ion{Ca}{2} near-infrared triplet.  The exact fitting region was
tailored for each galaxy to fit within a single echelle order and to
exclude emission lines such as [\ion{N}{1}] $\lambda5200$ and
[\ion{Fe}{7}] $\lambda5158$.  The measurement uncertainty was taken to
be the sum in quadrature of the fitting uncertainty from the
best-fitting template and the standard deviation of results from all
templates.  A minimum of six template stars were used for each
measurement.  We note that the problem of template mismatch in the
Mg/Fe abundance ratio, which is often noticeable in high-dispersion
galaxies \citep{bhs02}, does not significantly affect this sample.

\begin{figure}[t!]
\begin{center}
\scalebox{0.4}{\includegraphics{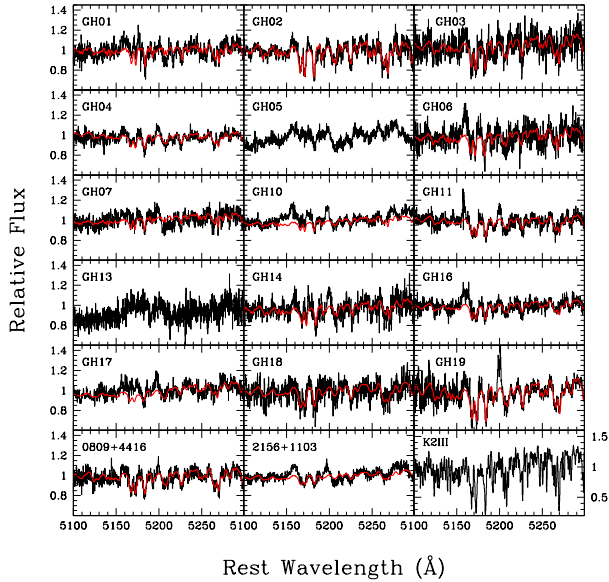}}
\end{center}
\caption{ESI observations of the spectral region around \mgb\ and
  Fe5270.  The spectra are flux-calibrated in $f_\lambda$ units and
  have been normalized to a median flux level of unity in order to
  show the varying amounts of nonstellar continuum dilution in the
  sample.  The red overplotted spectra are the best-fitting broadened
  template stars.  The spectrum of the K2III star HD 332344 is shown
  at the bottom right; note the expanded vertical scale of this panel.
\label{mgb}
}
\end{figure}

We were able to determine velocity dispersions for 13 of the 15
objects we observed from the GH sample, as well as the two new SDSS
objects, using the \mgb\ spectral region.  The other two galaxies were
dominated by AGN emission and no useful results could be recovered
from the template fitting.  For 7 galaxies we were able to measure
$\sigma$ from the \ion{Ca}{2} triplet region as well; the results are
generally consistent with the \mgb\ results but have larger
uncertainties because of the smaller spectral region used for the
fitting.  Thus, we use the \mgb\ measurements as the best estimates of
the stellar velocity dispersions.  The fitting results are listed in
Table \ref{table1} and illustrated in Figure \ref{mgb}.  (The figure
displays, for clarity, only a portion of the fitting region.)

We are unable to determine whether the measured velocity dispersions
are dominated by light from bulges, disks, or nuclear star clusters,
since in most cases the galaxies are only marginally resolved in the
SDSS images.  High-resolution imaging with the \emph{Hubble Space
Telescope} (\hst) would clarify the host galaxy morphologies and would
also make it possible to examine the behavior of the
$\mbh-L_{\mathrm{bulge}}$ relation for this sample.

\subsection{Black Hole Masses}

Black hole masses were derived by GH from the SDSS spectra using the
empirical relations from \citet{kas00}.  These relations combine the
broad-line region (BLR) radius (derived from the nonstellar 5100 \AA\
continuum luminosity) and broad-line width to obtain a virial estimate
of the central mass.  Single-epoch virial estimates of \mbh\ based on
this technique are thought to have a random scatter of roughly a
factor of 3, based on the consistency of the masses with the \msigma\
relation and other tests \citep[e.g.,][]{geb00b, fer01, ves02, nel04,
onk04}.  While FWHM(\hbeta) is most commonly used as a measure of the
velocity distribution of the BLR, the \hbeta\ emission lines in the
SDSS spectra have low S/N and GH used FWHM(\hal) as a surrogate.

The high S/N and spectral resolution of the ESI data make it possible
to perform decompositions of the \hal+[\ion{N}{2}] lines that are more
accurate than the previous measurements from SDSS spectra.  Following
the methods used by GH, we fit \hal+[\ion{N}{2}] using a
multi-Gaussian model.  All three narrow components (\hal\ and
[\ion{N}{2}] $\lambda\lambda6548, 6583$) were constrained to have the
same velocity width, and the wavelength separation and intensity ratio
of the [\ion{N}{2}] lines were fixed at their laboratory values.  Two
Gaussians were required to fit the broad component of \hal\ adequately
for most galaxies.  In most cases our results for FWHM(\hal) are
similar to those of GH, but in a few cases such as GH11 the ESI
spectrum yields a substantially larger width, which we attribute to
the difficulty of deblending the line profiles in SDSS spectra having
weak broad components and only marginally resolved narrow lines.  The
difference in linewidths could also be due in part to intrinsic
variability.  The statistical uncertainties on FWHM(\hal) from the
profile fits are small, typically only $\sim5-20$ \kms, while the true
uncertainties on the linewidths are dominated by systematics including
starlight subtraction errors, and we estimate these uncertainties to
be of order 10\% in FWHM(\hal).  We also measured the \hbeta\
linewidths using a double-Gaussian model to represent the broad
component.  For three objects the broad \hbeta\ line did not have
sufficient S/N to obtain a meaningful fit, and in one object (GH11)
broad \hbeta\ was not detected at all.  We do not find any systematic
trend for \hbeta\ to be significantly broader than \hal, and we use
the \hal\ lines to determine \mbh\ because their higher S/N enabled us
to measure the linewidths in a consistent manner for the entire
sample.

Since the Keck spectra were observed and extracted through a small
aperture and the slit losses are unknown, we continue to use the GH
measurements of the AGN continuum luminosity obtained from the SDSS
spectra.  These may slightly overestimate the true AGN luminosity
because of residual starlight in the SDSS spectra.  \hst\ $V$-band
imaging of this sample would be the best way to obtain more accurate
luminosities.

Single-epoch virial mass estimates for black holes in AGNs are subject
to potentially significant systematics, due to uncertainties in both
the slope $\gamma$ of the BLR radius-luminosity relation ($\rblr
\propto L^\gamma$) and the normalization factor $f$ in the virial
relation $\mbh = fr_{\mathrm{BLR}}v_{\mathrm{FWHM}}^2/G$.  The
radius-luminosity relation is not well determined for $\lambda
L_\lambda < 10^{43.5}$ erg s\per, and this presents particular
problems for our sample since it is necessary to extrapolate this
relation to luminosities as low as $\lambda L_\lambda < 10^{41.7}$ erg
s\per.  Following \citet{kas00} and \citet{ves02} we use $\gamma=0.7$;
setting $\gamma = 0.5$ instead \citep[e.g.,][]{shi03} would raise the
masses by 0.3 dex on average.  To obtain virial masses, we use the
recent calibration of $f$ from \citet{onk04} and \citet{pet04}, which
was derived by scaling the black hole masses of reverberation-mapped
Seyferts to fit the \msigma\ relation.  This calibration increases the
black hole masses by 0.27 dex relative to masses derived for the
assumption of isotropic orbits of BLR clouds \citep[e.g.,][]{kas00},
and shifts the sample to systematically higher masses than those
originally given by GH.  The revised masses are listed in Table 1.

\section{Discussion}

\begin{figure}
\begin{center}
\scalebox{0.4}{\includegraphics{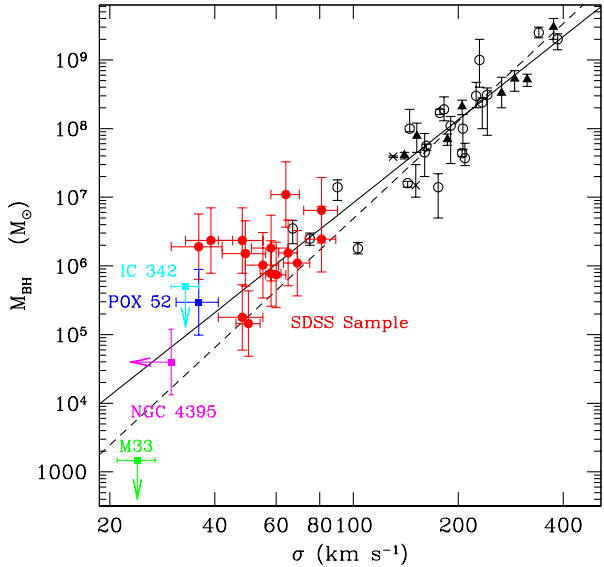}}
\end{center}
\caption{The \msigma\ relation for massive black holes in galaxy
  nuclei.  Points in black are black holes with dynamical mass
  measurements compiled by \citet{tre02}.  Open circles represent
  stellar-dynamical measurements, filled triangles are gas-dynamical
  detections, and crosses are from H$_2$O maser observations.  Red
  filled circles are the SDSS Seyfert 1 galaxies described in this
  paper.  Filled squares are measurements or upper limits for IC 342
  \citep{bok99}, POX 52 \citep{bar04}, NGC 4395 \citep{fh03}, and M33
  \citep{geb01}.  For the SDSS sample, POX 52, and NGC 4395, the
  vertical error bars correspond to a factor of 3 uncertainty in
  mass. The mass for POX 52 has been scaled upward by 0.27 dex from
  the value given by \citet{bar04} to reflect the new virial mass
  calibration from \citet{onk04}.  The solid and dashed lines
  represent the \msigma\ relations derived by \citet{tre02} and
  \citet{mf01}, respectively.
\label{msigma}
}
\end{figure}

Figure \ref{msigma} displays the \msigma\ relation for AGNs with low
black hole masses as well as for nearby galaxies with direct dynamical
measurements of \mbh.  For the SDSS sample, NGC 4395, and POX 52, the
vertical error bars represent a factor of 3 uncertainty in \mbh,
corresponding to the expected level of random scatter for single-epoch
measurements.  Overall, the SDSS AGNs lie close to the \msigma\
relation extrapolated to low masses.  Compared to the \citet{tre02}
\msigma\ relation with power-law slope 4.02, the SDSS galaxies have a
mean offset of 0.23 dex in \mbh\ and an rms scatter of 0.56 dex.  The
mean offset and rms scatter are 0.64 dex and 0.85 dex, respectively,
relative to the steeper \msigma\ relation of \citet{mf01} with slope
4.72.  (The mean offset from the Tremaine \etal\ \msigma\ relation
would be $-0.04$ dex if we used the virial relations from Kaspi \etal\
2000 rather than the Onken \etal\ calibration.)  The scatter of 0.56
dex relative to the shallower \msigma\ relation is only slightly
larger than the scatter of 0.4--0.5 dex for AGNs with
reverberation-based black hole masses \citep{onk04, nel04}.  This
overall consistency suggests that both the BLR radius-luminosity
relation and the \msigma\ relation can reasonably be extrapolated to
black holes in the range $\sim10^5-10^6$ \msun, well below the mass
scales in which these correlations have been calibrated.

Given the systematic uncertainties involved in deriving single-epoch
virial masses for these objects and the substantial scatter, we do not
attempt to fit the slope of the \msigma\ relation based on the new
data.  Nevertheless, our results may be useful as a constraint on
models making specific predictions for the low-mass end of the
\msigma\ relation.  For example, the hierarchical black hole growth
scenario of \citet{gra04} predicts that the \msigma\ slope should
substantially steepen below $\sigma = 150$ \kms\ as supernova feedback
becomes more efficient at slowing the AGN fueling rate.  The low-mass
AGN sample does not show such behavior.

We cannot rule out the possibility that the GH sample is only the
upper envelope of the population of black holes in low-mass galaxies.
The selection technique used by GH is only possible for broad-lined
AGNs, and the probability of detecting a galaxy as an AGN in the SDSS
archive depends on the AGN luminosity, which in turn depends on \mbh.
This potentially biases the sample towards preferential detection of
AGNs with the highest-mass black holes for a given $\sigma$.  There
are still few direct constraints on the masses of black holes in
inactive dwarf ellipticals or late-type spirals, and the case of M33
demonstrates that some late-type galaxies may not contain a black hole
at all \citep{geb01}.  Dynamical searches for low-mass black holes
should be continued whenever feasible in order to obtain an unbiased
sample of detections or upper limits in this mass range.

Based on their broad-line widths of $<2000$ \kms, the GH objects can
be classified as narrow-line Seyfert 1 galaxies (NLS1s).  Our
observations provide new data to test recent claims regarding the
\msigma\ relation of NLS1s.  Using [\ion{O}{3}] emission-line velocity
dispersions as a surrogate for stellar velocity dispersions,
\citet{gm04} found that \emph{ROSAT}-selected NLS1s lie below the
\msigma\ relation, typically by an order of magnitude, and concluded
that NLS1s are ``young'' AGNs with undermassive black holes.  We find,
on the contrary, that the low-mass NLS1 galaxies in the GH sample do
not deviate significantly from the local \msigma\ relation.  The use
of [\ion{O}{3}] emission-line velocity dispersions by \citet{gm04} in
place of stellar velocity dispersions may be the cause of this
discrepancy. As recently shown by \citet{bot04a}, NLS1s apparently
deviate from the \citet{nw96} empirical trend for Seyferts to have
similar stellar and emission-line velocity dispersions.  We will
examine the relation between the stellar and emission-line velocity
dispersions for the GH sample in a future paper, in the context of a
larger sample of Seyfert galaxies covering a wide range of black hole
masses.

~
~
~
~
~
~

\section{Conclusions}

We have measured the stellar velocity dispersions of 15 Seyfert 1
galaxies selected from the SDSS to have low-mass black holes based on
their broad-line widths and nonstellar continuum luminosities.  The
velocity dispersions for this sample are small, ranging from 36 to 81
\kms, and the results are consistent with a simple extrapolation of
the \msigma\ relation to low masses.  This initial view of the
demographics of low-mass black holes in AGNs implies a remarkable
continuity in the growth and evolution of black holes and their host
galaxies for \mbh\ ranging from $\sim10^5$ to over $10^9$ \msun.
Further refinement of the black hole mass estimates is possible if
reverberation mapping can be successfully carried out for AGNs having
$\mbh < 10^6$ \msun; this may be the best prospect for a definitive
measurement of the \msigma\ slope for low-mass black holes.

\acknowledgments 

We are grateful to George Djorgovski for obtaining data for this
project during the November 2003 Keck observing run.  We thank the
referee, Brad Peterson, for helpful comments that improved this paper.
Research by A.J.B. was supported by the UC Irvine Physical Sciences
Innovation Fund and by NASA through Hubble Fellowship grant
\#HST-HF-01134.01-A awarded by STScI.  Data presented herein were
obtained at the W.M. Keck Observatory, which is operated as a
scientific partnership among Caltech, the University of California,
and NASA. The Observatory was made possible by the generous financial
support of the W.M. Keck Foundation.  The authors wish to recognize
and acknowledge the very significant cultural role and reverence that
the summit of Mauna Kea has always had within the indigenous Hawaiian
community.

\end{document}